\documentclass[12pt]{iopart}
\usepackage{dcolumn,graphicx,color,booktabs,microtype,afterpage}
\usepackage{float}
\usepackage{amssymb}
\usepackage{url}  
\usepackage[charter,greekuppercase=italicized]{mathdesign}
\usepackage{sidecap}
\usepackage[mathlines]{lineno}

\usepackage{color}
\usepackage[colorlinks,plainpages=false,linkcolor=blue,urlcolor=blue,citecolor=blue,pdfpagemode=UseNone,pdfstartview=FitBH]{hyperref}
\graphicspath{{./}{figure/}}
\newcommand{\tcr}[1]{\textcolor{black}{#1}}

\newcommand{\LBP}{La$_{4}$Be$_{33}$Pt$_{16}$}

\begin{document}
	
	\makeatletter\renewcommand{\ps@plain}{%
		\def\@evenhead{\hfill\itshape\rightmark}%
		\def\@oddhead{\itshape\leftmark\hfill}%
		\renewcommand{\@evenfoot}{\hfill\small{--~\thepage~--}\hfill}%
		\renewcommand{\@oddfoot}{\hfill\small{--~\thepage~--}\hfill}%
	}\makeatother\pagestyle{plain}

\title[Probing the superconducting pairing of the La$_{4}$Be$_{33}$Pt$_{16}$]{Probing the superconducting pairing of the La$_{4}$Be$_{33}$Pt$_{16}$ alloy \tcr{via} muon-spin spectroscopy} 
\author{Tian\ Shang$^{1,2,*}$, Eteri\ Svanidze$^3$, Toni\ Shiroka$^{4,5,*}$}
\address{$^1$Key Laboratory of Polar Materials and Devices (MOE), School of Physics and Electronic Science, East China Normal University, Shanghai 200241, China}
\address{$^2$Chongqing Key Laboratory of Precision Optics, Chongqing Institute of East China Normal University, Chongqing 401120, China}
\address{$^3$Max Planck Institute for Chemical Physics of Solids, D-01187 Dresden, Germany}
\address{$^4$Laboratory for Muon-Spin Spectroscopy, Paul Scherrer Institut, CH-5232 Villigen PSI, Switzerland}
\address{$^5$Laboratorium f\"ur Festk\"orperphysik, ETH-H\"onggerberg, CH-8093 Z\"urich, Switzerland}

\eads{\mailto{tshang@phy.ecnu.edu.cn}, \mailto{tshiroka@phys.ethz.ch}}	
	
\date{\today}
	
\begin{abstract}
We report a study of the superconducting pairing of the
noncentrosymmetric \LBP\ alloy using muon-spin rotation and relaxation ({\textmu}SR) technique. 
Below $T_c = 2.4$\,K, \LBP exhibits bulk superconductivity (SC), 
	here characterized by heat-capacity and magnetic-susceptibility measurements.
	The temperature dependence of the superfluid density $\rho_\mathrm{sc}(T)$,
	extracted from the transverse-field {\textmu}SR measurements, reveals a nodeless SC in \LBP. The best fit of $\rho_\mathrm{sc}(T)$ using an $s$-wave model yields a magnetic penetration depth $\lambda_0 = 542$\,nm and a superconducting gap $\Delta_0 = 0.37$\,meV at zero Kelvin. 
	The single-gapped superconducting state 
	is further evidenced by the temperature-dependent electronic specific
	heat $C_\mathrm{e}(T)/T$ and the linear field-dependent electronic
	specific-heat coefficient $\gamma_\mathrm{H}(H)$. 
	The zero-field {\textmu}SR spectra collected in the normal- and superconducting states of \LBP\ are almost identical, confirming the absence of an additional field-related relaxation and, thus, of spontaneous magnetic fields below $T_c$. The nodeless SC combined with a preserved time-reversal symmetry in the superconducting state prove that the
	spin-singlet pairing is dominant in \LBP. \tcr{This material represents yet another example of a complex
	system showing only a conventional behavior, in spite of a
	noncentrosymmetric structure and a sizeable spin-orbit coupling.}
\end{abstract}
		
%
%

%

\maketitle
%
%

	
\section{Introduction}

Unconventional superconductivity (SC) is one of the trending
topics in condensed-matter physics of the last decades~\cite{Sigrist1991}. The superconductors whose crystal structure lacks an inversion center are known as  noncentrosymmetric superconductors (NCSCs). In NCSCs, the broken inversion symmetry allows for an antisymmetric spin-orbit coupling (ASOC), which might lead to mixed spin-singlet and spin-triplet pairings~\cite{Bauer2012}. The degree of such mixing is believed to be closely related to the strength of ASOC~\cite{Bauer2012,Yip2014}.
Since the discovery of unconventional SC in the first
noncentrosymmetric CePt$_3$Si heavy-fermion compound~\cite{Bauer2004},
a series of NCSCs have been discovered and investigated~\cite{Bauer2012,Smidman2017,Ghosh2020b}. Owing to the mixed pairing, some of them 
exhibit several exotic superconducting properties, such as, upper critical fields beyond the Pauli limit~\cite{Bauer2004,Carnicom2018,Su2021,Balakirev2015,Bao2015,Tang2017,Tang2015}, nodes in the superconducting  gap~\cite{yuan2006,nishiyama2007,bonalde2005CePt3Si,Shang2020}, multigap SC~\cite{kuroiwa2008}, as well as time-reversal symmetry (TRS) breaking in the superconducting state~\cite{Shang2020,Hillier2009,Barker2015,Shang2020b,Singh2014,Shang2018a,Shang2018b,Shang2020ReMo,Shang2022b,Shang2022c,Shang2021a}.
Very recently, NCSCs have also become a fertile platform to search for and investigate the Majorana zero modes and topological SC, with potential applications to quantum computation~\cite{Sato2017,Qi2011,Kallin2016,Kim2018,Ali2014}. \tcr{For instance, topological surface states have been found in PbTaSe$_2$~\cite{Guan2016,Bian2016}, $\beta$-Bi$_2$Pd~\cite{Sakano2015}, and $\alpha$-BiPd~\cite{Sun2015,Neupane2016} NCSCs}.

\tcr{Only a few families of NCSCs have been found to break the TRS below $T_c$. These include 
$\alpha$-Mn-type Re$T$ ($T$ = transition metal)~\cite{Singh2014,Shang2018a}, Th$_7$Fe$_3$-type La$_7$T$_3$~\cite{Barker2015,Singh2020LaRh,Arushi2021}, CeNiC$_2$-type LaNiC$_2$~\cite{Hillier2009},
TiFeSi-type $T$RuSi~\cite{Shang2022c}, V$_3$S-type Zr$_3$Ir~\cite{Shang2020b}, and ThSi$_2$-type La$T$Si and CaPtAs~\cite{Shang2022c,Shang2020}. 
Most of the above cases exhibit nodeless superconductivity, while CaPtAs represents a rare case, which accommodates both broken TRS and nodal SC~\cite{Shang2020}. 
Some of these NCSCs exhibit a significant ASOC (e.g., CaPtAs, LaPtGe), while in other cases the SOC is rather weak or even can be ignored (e.g., LaNiC$_2$, LaNiSi).  
Later on, muon-spin spectroscopy studies of elementary rhenium and Re-Mo alloys revealed that the TRS breaking appears only in Re-rich compounds, and it is 
independent of the centro- or noncentrosymmetric crystal structure~\cite{Shang2020ReMo,Shang2018b,Shang2021a}.   
In general, the causes behind TRS breaking in NCSCs are not yet fully understood and remain an intriguing open question.
}	

Up to now, despite the many NCSCs that have been discovered,
only a few of them are known to host mixed spin-singlet and spin-triplet pairings.
Notable examples include Pt- or Pd compounds, such as Li$_2$Pt$_3$B and Li$_2$Pd$_3$B. 
While Li$_2$Pd$_3$B behaves as an $s$-wave superconductor, characterized by a fully-gapped superconducting state~\cite{yuan2006}, the enhanced ASOC due to the Pt-for-Pd substitution makes Li$_2$Pt$_3$B a nodal superconductor that exhibits typical features of spin-triplet pairing~\cite{yuan2006,nishiyama2007}, e.g.,
its spin susceptibility does not change upon crossing $T_c$~\cite{nishiyama2007}. 
\tcr{Unfortunately, single crystals of the Li$_2$(Pd,Pt)$_3$B family have not been grown to date, while grain-boundary effects could not be excluded in previous studies of polycrystalline samples.
Similarly, in $\alpha$-BiPd NCSC, unconventional pairing states have been identified via the Little-Parks effect and point-contact Andreev reflection~\cite{Mondal2012,Xu2020}. Yet, muon-spin spectroscopy studies reveal  
its conventional nature, i.e., a fully-gapped superconducting state with a preserved TRS~\cite{Ramakrishnan2017,Parzyk2014}.}
Another representative example is CaPtAs, whose ASOC is also significant~\cite{Xie2019}.
Below $T_c$, it exhibits the breaking of TRS, as well as nodes in the superconducting gap~\cite{Shang2020,Xie2019}.  
Although many NCSCs show a relatively large ASOC,
they still behave as conventional superconductors, characterized by a nodeless SC with preserved TRS, typical features of spin-singlet pairing~\cite{Shang2021b,Sharma2020,Shang2023,Tay2023,Ptok2019}.
Clearly, it is of great interest to search for other NCSCs which exhibit triplet pairing or breaking of the TRS and, thus, behave as unconventional- or topological superconductors.	

In this respect, \LBP\ represents a promising candidate material, since both La and Pt are heavy metals which might promote strong ASOC effects.
It adopts a cubic Th$_{4}$Be$_{33}$Pt$_{16}$-like crystal structure and belongs to a new family of noncentrosymmetric superconductors, $RE$$_4$Be$_{33}$Pt$_{16}$ ($RE$ = Y, La-Nd, Sm-Lu, Th, U), that form across the lanthanide series~\cite{Svanidze2021,Amon2020,Kozelj2021}. 
Their diverse physical properties encompass anti- and ferromagnets, paramagnets, and superconductors.
This diversity is most likely related to the nature of rare-earth elements and to the complex multicentre interactions within the Be-Pt framework. 
Given its 212(!) atoms per unit cell, the Th$_{4}$Be$_{33}$Pt$_{16}$-type superconducting family is one of the most complex NCSCs discovered to date.
Although certain superconducting properties of \LBP\ have been previously studied using macroscopic techniques~\cite{Svanidze2021}, its superconducting pairing remains largely unexplored at a microscopic level. 

In this paper, we investigate the SC of noncentrosymmetric \LBP\ alloy using mostly the muon-spin rotation and relaxation ({\textmu}SR) technique. 
Our {\textmu}SR results demonstrate that, despite a noncentrosymmetric crystal structure and a sizeable SOC, \LBP\ exhibits a fully-gapped SC with preserved TRS.

\section{Experimental details}

\begin{figure}[thb]
	\centering
	\includegraphics[width=0.5\textwidth]{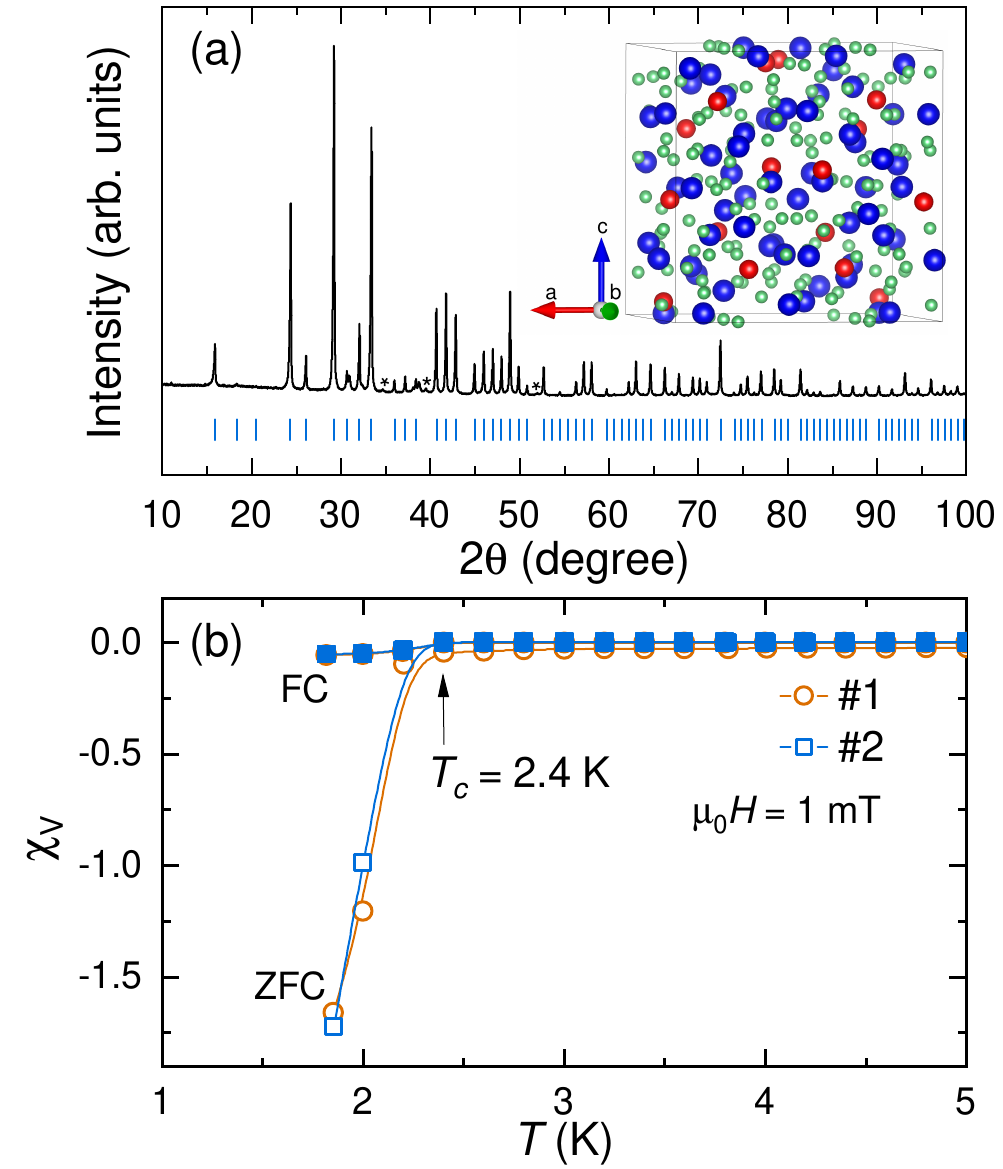} 
	\caption{\label{fig:XRD}\tcr{(a) Powder XRD pattern of the \LBP\ alloy.
		The calculated Bragg peak positions using the noncentrosymmetric
		cubic $I\bar{4}3d$ space group are presented below the pattern.
		The inset shows the crystal structure (unit cell).
		Red, green, and blue spheres represent La, Be, and Pt atoms, respectively.
		(b) The magnetic susceptibility $\chi_\mathrm{V}(T)$ as a function of temperature for two different batches of the \LBP\ alloy (marked as \#1 and \#2).
		Both the ZFC- and FC $\chi_\mathrm{V}(T)$ were measured in a magnetic field of $\mu_0H$ = 1\,mT. 
		The well-distinct ZFC- and FC curves indicate a strong-pinning type-II SC in \LBP,
		as confirmed also by TF-{\textmu}SR measurements (see below).}}
\end{figure}

Since pure Be metal is toxic, the \LBP\ samples were prepared in an argon-filled glove box (MBraun, $p$(H$_2$O/O$_2$)$ < 0.1$\,ppm) in a specialized laboratory. Polycrystalline \LBP\ samples were prepared by arc melting
La (Ames Laboratory, $>$99.99\%), Pt (ChemPUR, $>$99.9\%), and Be (Heraeus, $>$99.9\%) metals in a ratio of 10:30:60. Different batches of \LBP\ samples were prepared and characterized. To improve the homogeneity, all samples were flipped and re-melted several times. The mass loss was found to be less than 2\% for the final as-grown samples. 
The details of the sample synthesis can be found also in Ref.~\cite{Svanidze2021}.
Powder x-ray diffraction (XRD) measurements were performed at room temperature on a Huber G670
image-plate Guinier camera with a Ge monochromator (Cu K$\alpha_1$, $\lambda = 1.54056$\,\AA).

Heat-capacity and magnetic susceptibility measurements were carried out on a Quantum Design physical property measurement system (PPMS) and a magnetic property measurement system (MPMS), respectively. The {\textmu}SR measurements were performed at the multipurpose surface-muon spectrometer 
(Dolly) at the $\pi$E1 beamline of the Swiss muon source at Paul 
Scherrer Institut (PSI), Villigen, Switzerland.
The \LBP\ alloy samples were cut into few thin plates (with a thickness of $\sim$1.5\,mm) and then
mounted on a 25-{\textmu}m thick copper foil to cover an area 6--8\,mm in diameter.
The time-differential {\textmu}SR spectra comprised both transverse-field (TF)
and zero-field (ZF) {\textmu}SR measurements, performed upon heating the sample. The {\textmu}SR data were analyzed by means of the \texttt{musrfit} software package~\cite{Suter2012}.

\section{Results and discussion} 
We first checked the phase purity and crystal structure of the synthesized \LBP\ alloy samples by performing XRD measurements at room temperature. The XRD pattern confirms a cubic crystal structure with a noncentrosymmetric $I\bar{4}3d$ space group for \LBP\ [Fig.~\ref{fig:XRD}(a)]. The crystal structure (unit cell) of \LBP\ is plotted in the inset of Fig.~\ref{fig:XRD}(a).
Further details on the crystal structure can be found in Ref.~\cite{Svanidze2021}. 
Since the evaporation losses of Be metal are unavoidable, small inclusions of secondary phases (with a weight of $<$9\%) were detected (indicated by the star symbols). 
According to the energy-dispersive x-ray (EDX) spectroscopy measurements, 
the secondary phases are most likely LaPt and LaPt$_2$ binary alloys, both of which remain paramagnetic in the studied temperature range (i.e., $\geq 0.3$\,K)
and, thus, their influence on the superconductivity of \LBP\ is negligible~\cite{Joseph1972,Holt1981}.

\begin{figure}[tbh]
	\centering
	\includegraphics[width=0.5\textwidth]{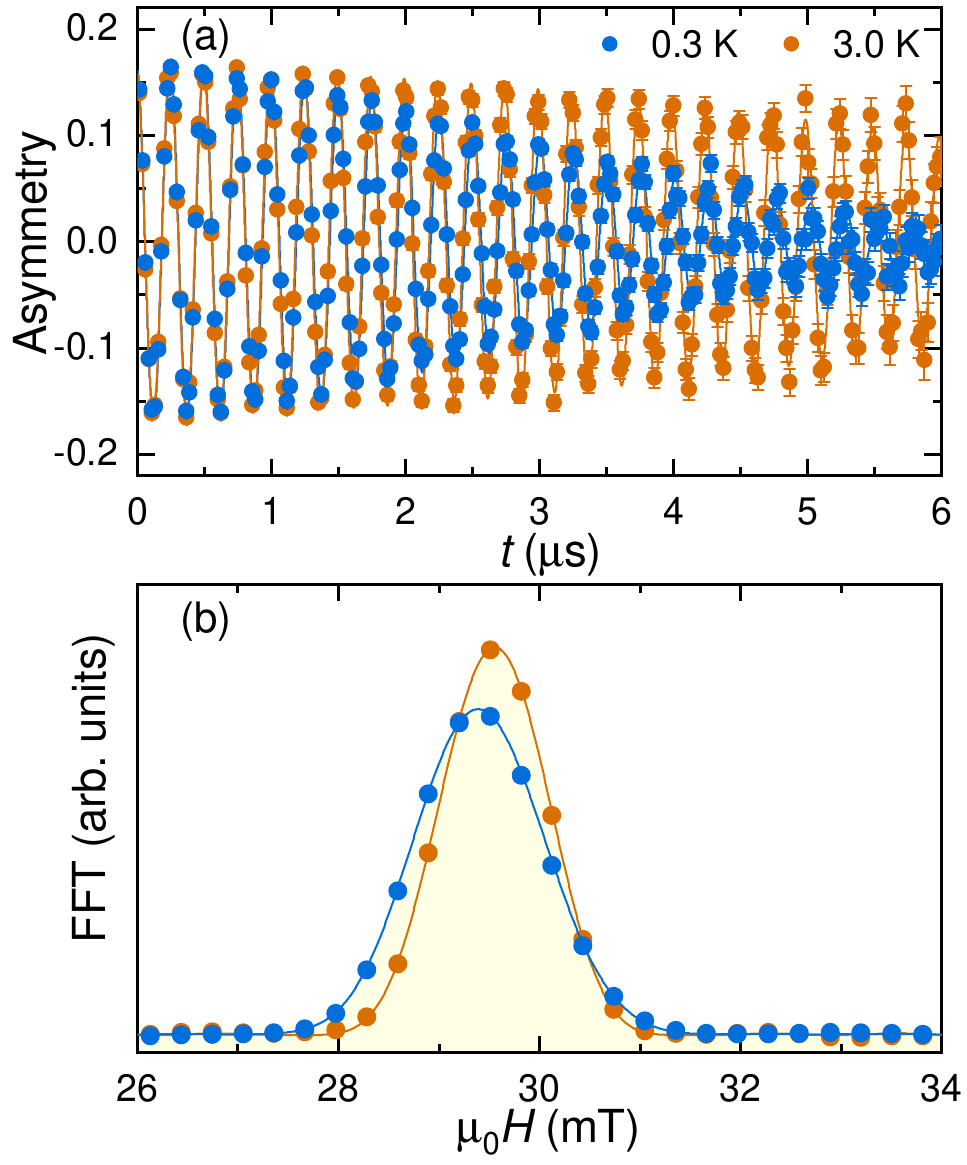}%
	\caption{\label{fig:TF-muSR}\tcr{(a) Superconducting- (0.3\,K) and normal-state (3.0\,K) 
			TF-$\mu$SR spectra for \LBP\ show clear differences in relaxation. 
			A magnetic field of 30\,mT was applied in the normal state and the $\mu$SR spectra were collected upon heating the sample.
		(b) Fast Fourier transforms FFT of the TF-$\mu$SR spectra shown in panel (a) as a function of magnetic field.  
		Solid lines through the data are fits to Eq.~\ref{eq:TF_muSR}.
		Note the broadening of the Gaussian field distribution in the superconducting state due to the formation of FLL.}}
\end{figure}

The magnetic susceptibility of the \LBP\ alloys was measured using both zero-field-cooling (ZFC) and field-cooling (FC) protocols in an applied magnetic field of 1\,mT.
A clear diamagnetic susceptibility appears below the onset of superconducting transition
at $T_c$ = 2.4\,K [see Fig.~\ref{fig:XRD}(b)].
The two different batches of 
samples show almost identical $T_c$ values and superconducting volume fractions. 
Since the samples (\#1 and \#2) used for magnetic susceptibility measurements have irregular geometries, their demagnetization factors are rather difficult to estimate. 
In any case, a large diamagnetic response (i.e., $\chi_\mathrm{V} > -1.5$) 
indicates bulk SC in \LBP. This result was further confirmed by TF-{\textmu}SR and heat-capacity measurements (see below).

\begin{figure}[tbh]
	\centering
	\vspace{-2mm}
	\includegraphics[width=0.5\textwidth]{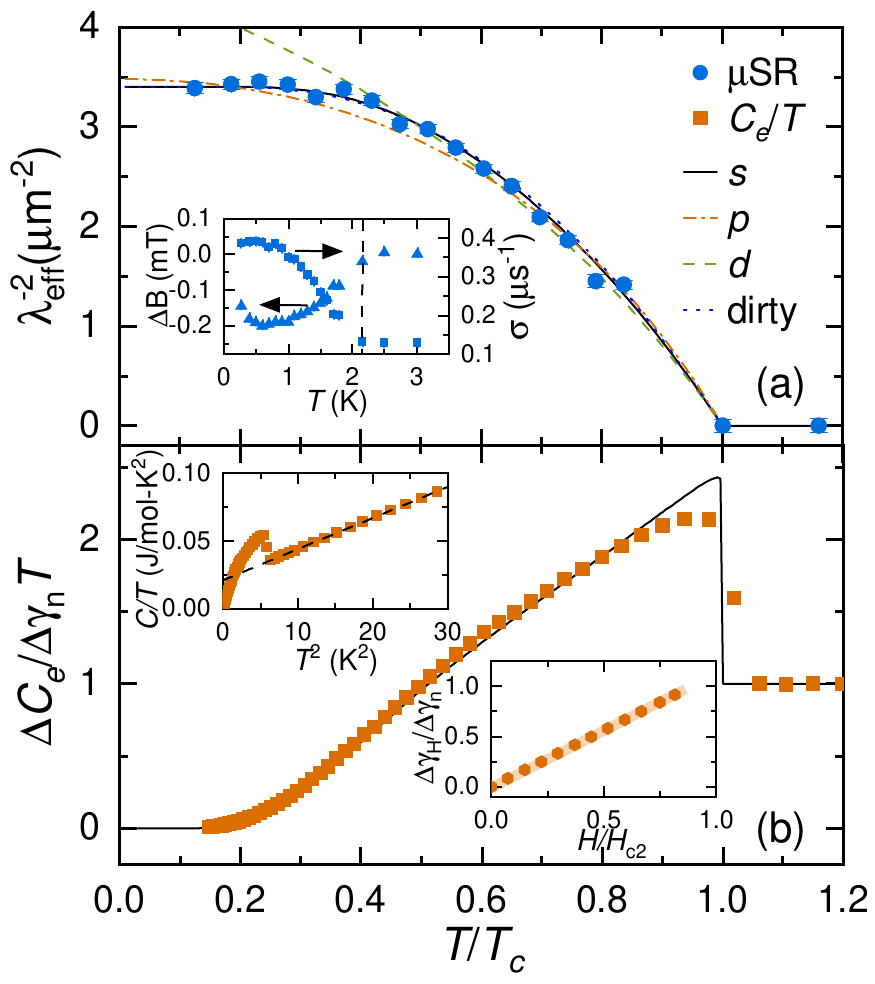}%
	\caption{\label{fig:rhos}\tcr{Temperature-dependent superfluid density $\rho_\mathrm{sc}(T)$ [$\propto$$\lambda_\mathrm{eff}^{-2}(T)$] (a) and normalized zero-field electronic specific heat $C_\mathrm{e}(T)/\gamma_\mathrm{n}T$ (b) for \LBP. The various line types
			represent fits to Eq.~\ref{eq:rhos} using $s$-, $p$-, and $d$-wave models. The dotted line is a fit to the BCS equation in the dirty limit. 
		    The muon-spin relaxation rate $\sigma(T)$ (right axis) and diamagnetic shift $\Delta$$B(T)$ (left axis) are shown in the inset of panel (a) as a function of temperature.
	    	The upper inset in (b) depicts the raw $C$/$T$ data versus $T^2$. The phonon contribution was estimated by fitting the specific heat to $C$/$T = \gamma_\mathrm{n}  + \beta T^2$ (dashed line). 
	        The specific-heat coefficient under various magnetic fields $\Delta\gamma_\mathrm{H}$ normalized to $\Delta\gamma_\mathrm{n}$ is
			plotted in the bottom inset of panel (b) vs. the reduced magnetic field $H/H_\mathrm{c2}(0)$.
			For each applied magnetic field, $\Delta\gamma_\mathrm{H}$ is obtained as the linearly extrapolated electronic specific heat $C_\mathrm{e}$/$T$ vs $T^2$ in the superconducting state to zero temperature, to which we then subtract $\gamma_\mathrm{res}$.
	     	The solid black line in (b) is a fit to Eq.~\ref{eq:entropy} using a fully-gapped $s$-wave model. The specific-heat data under various magnetic fields were adopted from Ref.~\cite{Svanidze2021}.}}
\end{figure}

To further investigate the gap symmetry and superconducting pairing of \LBP,
we carried out systematic TF-{\textmu}SR measurements in an applied field of 30\,mT at various temperatures. 
For the TF-{\textmu}SR measurements, the magnetic field is applied perpendicular to the muon-spin direction, leading the precession of the muon spin.
Figure~\ref{fig:TF-muSR}(a) plots the representative superconducting- and 
	normal-state TF-{\textmu}SR spectra for \LBP. The enhanced muon-spin relaxation (also known as damping) in the superconducting state is clealry visible. Such an enhanced damping in the TF-{\textmu}SR spectra is due to the formation of a flux-line lattice (FLL) during the field-cooling process, which generates an inhomogeneous field distribution~\cite{Yaouanc2011}.
	The broadening of field distribution in the superconducting state is clearly reflected in the fast Fourier transforms (FFT) of the TF-$\mu$SR spectra [see Fig.~\ref{fig:TF-muSR}(b)]. 
	Since the field distribution is quite symmetric, the TF-{\textmu}SR spectra can be fitted using a single-oscillation model, described by~\cite{Maisuradze2009}:
\begin{equation}
	\label{eq:TF_muSR}
	A_\mathrm{TF}(t) = A \cos(\gamma_{\mu} B t + \varphi) e^{- \sigma^2 t^2/2} +
	A_\mathrm{bg} \cos(\gamma_{\mu} B_\mathrm{bg} t + \varphi).
\end{equation}
Here $A$, $A_\mathrm{bg}$ and $B$, $B_\mathrm{bg}$ stand for the initial asymmetries and internal fields 
of the sample and sample holder (i.e., Cu plate), respectively; $\sigma$ is the Gaussian relaxation rate, a measure of the field distribution; $\gamma_{\mu}$ and $\varphi$ are the muon gyromagnetic ratio and a shared initial phase, respectively. As shown in the inset of Fig.~\ref{fig:rhos}(a), in the normal state, the muon-spin relaxation rates $\sigma$ are small ($\sim$0.13 $\mu$s$^{-1}$) and almost temperature invariant. However, in the superconducting state, $\sigma$ starts to increase due to the formation of FLL below $T_c$, reaching $\sim$0.39 $\mu$s$^{-1}$ at 0.3\,K (i.e., at base temperature).
Simultaneously, a diamagnetic field shift $\Delta B = B_\mathrm{s} - B_\mathrm{appl}$ shows up below the onset of SC, where $B_\mathrm{appl}$ stands for the applied magnetic field (30\,mT, in our case).
Since the nuclear relaxation rate $\sigma_\mathrm{n}$ of \LBP\ is generally temperature independent in the covered temperature range and much smaller than the superconducting  relaxation rate $\sigma_\mathrm{sc}$, as evidenced by the ZF-{\textmu}SR measurements (see below), the latter can be extracted from the relation $\sigma_\mathrm{sc} = \sqrt{\sigma^2 - \sigma^{2}_\mathrm{n}}$. 

Considering that the transverse magnetic field we apply is much smaller than the upper critical field of \LBP, i.e., $B_\mathrm{appl}$/$B_\mathrm{c2}$ $\sim$ 0.02, when extracting the effective magnetic penetration depth $\lambda_\mathrm{eff}$ from the measured superconducting relaxation rate $\sigma_\mathrm{sc}$ we can ignore the effects of overlapping vortex cores~\cite{Svanidze2021}. Therefore, the $\lambda_\mathrm{eff}$ was calculated using $\sigma_\mathrm{sc}^2(T)/\gamma^2_{\mu} = 0.00371\Phi_0^2/\lambda_\mathrm{eff}^4(T)$~\cite{Barford1988,Brandt2003}, where $\Phi_0 = 2.07 \times 10^{3}$\,T~nm$^{2}$ is the quantum of magnetic flux.  	As shown in the main panel of Fig.~\ref{fig:rhos}(a), the temperature-dependent inverse square of magnetic penetration depth $\lambda_\mathrm{eff}^{-2}$ is summarized as a function of reduced temperature $T$/$T_c$ for \LBP. Since $\lambda_\mathrm{eff}^{-2}$ is proportional to the superfluid density $\rho_\mathrm{sc}(T)$, the temperature-dependent $\lambda_\mathrm{eff}^{-2}(T)$ reflects the superconducting pairing. 
The superfluid density $\rho_\mathrm{sc}(T)$ was analyzed by using	different models, which can be described by:
\begin{equation}
	\label{eq:rhos}
	\rho_\mathrm{sc}(T) = 1 + 2\, \Bigg{\langle} \int^{\infty}_{\Delta_\mathrm{k}} \frac{E}{\sqrt{E^2-\Delta_\mathrm{k}^2}} \frac{\partial f}{\partial E} \mathrm{d}E \Bigg{\rangle}_\mathrm{FS}.
\end{equation}
Here, $f = (1+e^{E/k_\mathrm{B}T})^{-1}$ represents the Fermi function~\cite{Tinkham1996}.
$\Delta_\mathrm{k}(T) = \Delta(T) \delta_\mathrm{k}$ is an angle-dependent
gap function, where $\Delta$ is the maximum gap value and $\delta_\mathrm{k}$ is the 
angular dependence of the gap. For the $s$-, $p$-, and $d$-wave model, $\delta_\mathrm{k}$ = 1, $\sin\theta$, and $\cos2\phi$, respectively, where $\theta$ and $\phi$ are the azimuthal angles.
The evolution of superconducting gap as a function of temperature is generally written as $\Delta(T) = \Delta_0 \mathrm{tanh} \{1.82[1.018(T_\mathrm{c}/T-1)]^{0.51} \}$, where $\Delta_0$ is the zero-temperature energy gap~\cite{Tinkham1996,Carrington2003}.
We applied three different models, including $s$-, $p$-, and $d$-wave, to analyze the $\rho_\mathrm{sc}(T)$ data of \LBP.
As shown by the dash-dotted line in Fig.~\ref{fig:rhos}(a), the $p$-wave model yields a zero-temperature superconducting gap $\Delta_0$ =  0.52(2)\,meV and a magnetic penetration depth $\lambda_\mathrm{0} =536(3)$\,nm. 
In the case of $d$-wave model (see dashed line), the gap size $\Delta_0 = 0.44(1)$\,meV is comparable to the value obtained from the $p$-wave model, but $\lambda_\mathrm{0}$ [470(3)\,nm] is much shorter. 
While for the $s$-wave model, a $\Delta_0$ = 0.37(1)\,meV and $\lambda_\mathrm{0} =542(3)$\,nm were obtained (see solid line).
As can be clearly seen in Fig.~\ref{fig:rhos}(a), the $d$-wave model agrees rather poorly with the experimental data when $T/T_c$ $\le$ 0.4. Similarly, the $p$-wave model fails to reproduce the superfluid density
 $\rho_\mathrm{sc}(T)$ at 
0.2 $\le$ $T/T_c$ $\le$ 0.6. In fact, the temperature-independent $\rho_\mathrm{sc}(T)$ below $T_c$/3 is more consistent with a fully-gapped superconducting state in \LBP.
Hence, only the $s$-wave model describes fairly well the experimental data across the full temperature range [see solid line in Fig.~\ref{fig:rhos}(a)]. 
We also applied the dirty-limit model to analyze the superfluid density. 
Within the BCS approximation, the temperature evolution of the superfluid density in the dirty limit can be written as 
$\rho_\mathrm{sc}(T) = \frac{\Delta(T)}{\Delta_0} \mathrm{tanh} \left[\frac{\Delta(T)}{2k_\mathrm{B}T}\right]$~\cite{Tinkham1996}, where $\Delta(T)$ is the same as in Eq.~\ref{eq:rhos}.
This model also describes quite well the $\rho_\mathrm{sc}(T)$ data, and yields a zero-temperature gap
$\Delta_0 =  0.32(1)$\,meV, the latter being comparable to the clean $s$-wave model. 

\begin{figure}[tbh]
	\centering
	\vspace{-2mm}
	\includegraphics[width=0.5\textwidth]{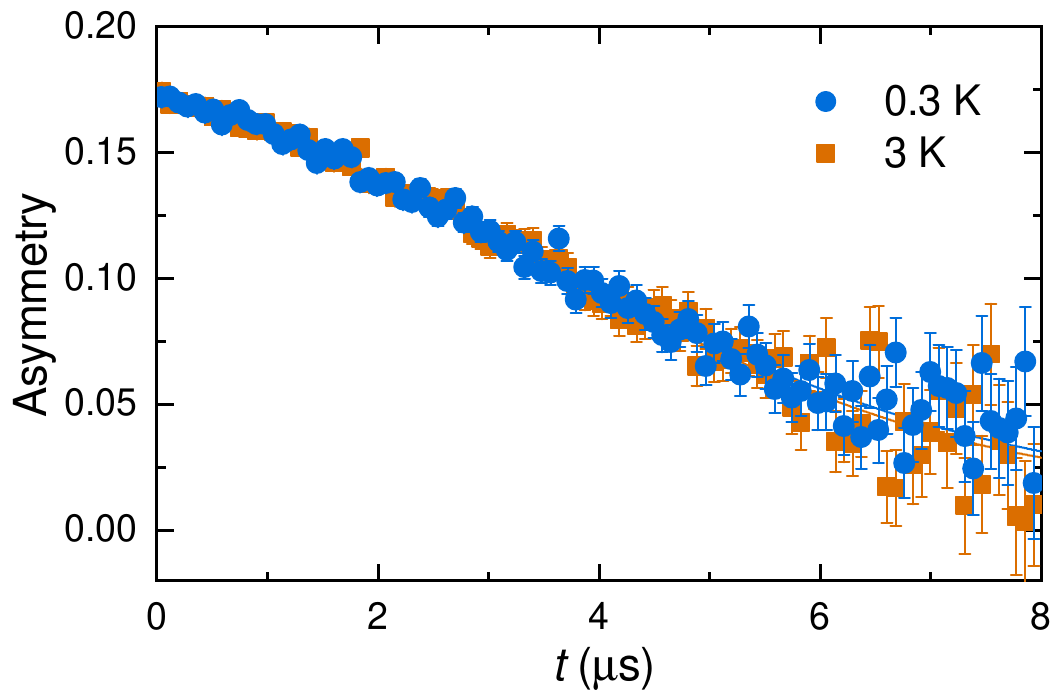}%
	\caption{\label{fig:ZF-muSR}Superconducting- (0.3\,K) and normal-state (3.0\,K) 
		ZF-{\textmu}SR spectra of the \LBP\ alloy. Both spectra almost overlap,
		indicating the absence of an additional muon-spin relaxation
		and, thus, the absence of TRS breaking in the superconducting state.}
\end{figure}

To confirm the above conclusion about nodeless SC in \LBP, we further analyzed the electronic specific
heat in the superconducting state. 
\tcr{The electronic specific heat $\Delta$$C_\mathrm{e}/T$ of \LBP\ was obtained by subtracting the phonon contribution $\beta$$T^2$ and that of the residual LaPt and LaPt$_2$ phases $\gamma_\mathrm{res}$}	
[see details in the upper inset of Fig.~\ref{fig:rhos}(b)] from the measured specific-heat data. \tcr{The estimated $\gamma_\mathrm{res}$ is relatively
small, here about 13\% of the normal-state electronic specific-heat coefficient $\gamma_\mathrm{n}$.}	
\tcr{Figure~\ref{fig:rhos}(b) shows the normalized electronic specific heat $\Delta$$C_\mathrm{e}/\Delta\gamma_\mathrm{n}T$ vs the reduced temperature $T/T_c$ for \LBP, where $\Delta$$C_\mathrm{e}/T = C_\mathrm{e}/T - \gamma_\mathrm{res}$ and $\Delta\gamma_\mathrm{n} = \gamma_\mathrm{n} - \gamma_\mathrm{res}$.} 
According to the BCS expression, the entropy of the superconducting state can be calculated as~\cite{Tinkham1996}:
\begin{equation}
	\label{eq:entropy}
	S(T) = -\frac{6\gamma_\mathrm{n}}{\pi^2 k_\mathrm{B}} \int^{\infty}_0 [f\mathrm{ln}f+(1-f)\mathrm{ln}(1-f)]\,\mathrm{d}\epsilon,
\end{equation}
where $f$ is again the Fermi function, as used in Eq.~\ref{eq:rhos}. 
Then, the electronic specific heat below $T_c$ can be derived according to $C_\mathrm{e}(T)/T = \frac{dS(T)}{dT}$.
Since the superfluid density $\rho_\mathrm{sc}(T)$ is almost temperature-independent below $\frac{1}{3}\,T_c$, a typical feature of nodeless SC, we used the $s$-wave model to analyze the electronic specific heat data of \LBP. 
Similar to the $\rho_\mathrm{sc}(T)$, the $s$-wave model also agrees very well with the $\Delta$$C_\mathrm{e}(T)/T$ data [see solid line in Fig.~\ref{fig:rhos}(b)], with $\gamma_\mathrm{n}$ = 21.0(1) \,mJ/mol-K$^2$ and a SC gap $\Delta_0 =  0.36(1)$\,meV. 
Such a gap value is highly consistent with the value obtained
from the TF-$\mu$SR results. 

The single-gap SC of \LBP\ can be further inferred from the specific-heat data under various magnetic fields, \tcr{here taken from Ref.~\cite{Svanidze2021}}.
As shown in the inset of Fig.~\ref{fig:rhos}(b), the electronic specific-heat coefficient \tcr{(after subtracting $\gamma_\mathrm{res}$)} determined under various magnetic fields, \tcr{ $\Delta\gamma_\mathrm{H}(H) [= \gamma_\mathrm{H}(H) - \gamma_\mathrm{res}$]}, is normalized to its zero-field normal-state value \tcr{$\Delta\gamma_\mathrm{n} (= \gamma_\mathrm{n} - \gamma_\mathrm{res}$)},
and then reported as a function of the reduced magnetic field $H/H_\mathrm{c2}(0)$.
In case of a nodal superconductor, $\gamma_\mathrm{H}(H)$ is usually expected to follow a square-root field dependence, i.e., $\sqrt{H}$~\cite{Volovik1993,Wen2004}.
While, in multigap superconductors, $\gamma_\mathrm{H}(H)$ shows a complex behavior, which depends strongly 
on the gap sizes and cannot be described with a simple formula~\cite{Chen2017,Bouquet2001a,Huang2007}. 
In this case, $\gamma_\mathrm{H}(H)$ changes slope at the field value that suppresses the smaller superconducting gap.
In contrast to the above cases, $\Delta\gamma_\mathrm{H}(H)$ of \LBP\ shows a purely linear field dependence. This is consistent with a fully-gapped superconducting state with a single energy gap~\cite{Caroli1964}, as observed also in other single-gap superconductors~\cite{Shang2018b,Shang2020NbIr}.

Due to the presence of spin-triplet pairing, certain NCSCs have been
found to break TRS in their superconducting state~\cite{Shang2020,Hillier2009,Barker2015,Shang2020b,Singh2014,Shang2018a,Shang2018b,Shang2020ReMo,Shang2022b}.
ZF-{\textmu}SR is one of the few very sensitive techniques that can
detect the weak spontaneous magnetic fields occurring in this case 
below the superconducting transition temperature.
Here, we also carried out the ZF-{\textmu}SR measurements to search for a possible breaking of TRS in \LBP.
The ZF-{\textmu}SR spectra in the normal- and superconducting states of \LBP\ are shown in Fig.~\ref{fig:ZF-muSR}.
The lack of fast decays and of coherent oscillations in the ZF-{\textmu}SR indicates the absence of magnetic fluctuations or magnetic order in \LBP.
Therefore, the muon-spin relaxation rates in \LBP\ is mostly determined by the randomly oriented nuclear moments, which normally are modeled by a Gaussian Kubo-Toyabe relaxation function
$G_\mathrm{KT} = [\frac{1}{3} + \frac{2}{3}(1 -\sigma_\mathrm{ZF}^{2}t^{2})\,\mathrm{e}^{-\sigma_\mathrm{ZF}^{2}t^{2}/2}]$~\cite{Kubo1967,Yaouanc2011}.
Here, $\sigma_\mathrm{ZF}$ represents the Gaussian relaxation rate in zero field, whose value is comparable to the $\sigma_\mathrm{n}$ determined from TF-{\textmu}SR measurements [see details in the inset of Fig.~\ref{fig:rhos}(a)]. 
As shown by the solid lines in Fig.~\ref{fig:ZF-muSR}, the ZF-$\mu$SR spectra of \LBP\
were fitted with $A_\mathrm{ZF}(t) = A G_\mathrm{KT} \mathrm{e}^{-\Lambda_\mathrm{ZF} t} + A_\mathrm{bg}$, where  $\Lambda_\mathrm{ZF}$ is a zero-field exponential relaxation rate that accounts for an additional electronic contribution. The obtained muon-spin relaxation rates are 
$\sigma_\mathrm{ZF} = 0.157(5)$\,$\mu$s$^{-1}$ and $\Lambda_\mathrm{ZF} = 0.069(7)$\,$\mu$s$^{-1}$ at 0.3\,K, while at 3\,K, the $\sigma_\mathrm{ZF} = 0.167(5)$\,$\mu$s$^{-1}$ and $\Lambda_\mathrm{ZF} = 0.057(7)$\,$\mu$s$^{-1}$.
Clearly, both the Gaussian- and the exponential relaxation rates are almost the same in the superconducting- and normal states of \LBP, with their differences lying within the standard deviations. 
Such identical {\textmu}SR relaxation rates unambiguously confirm a preserved TRS in the superconducting state of \LBP.
Hence, our ZF-{\textmu}SR results, combined with TF-{\textmu}SR, suggest a conventional nodeless SC with a preserved TRS in \LBP.

According to the electronic band-structure calculations, the density of states (DOS) is mostly dominated by the Pt-5$d$ and Be-2$p$ orbitals near the Fermi level $E_\mathrm{F}$,
while the contribution from La-5$d$ or 4$f$ orbitals is negligible~\cite{Svanidze2021}. 
Since the contribution to the DOS at the $E_\mathrm{F}$ of the Be-2$p$ orbitals is almost twice that of the Pt-5$d$ orbitals, it is not surprising that \LBP\ exhibits a small  
band splitting $E_\mathrm{SOC}$.
Hence, despite the presence of heavy elements that usually exhibit a large SOC, such as La and Pt, \LBP\ shows rather weak ASOC effects.
As a consequence, the inclusion of SOC produces negligible changes to
the valence states or DOS at $E_\mathrm{F}$. The exotic superconducting properties of NCSCs are generally believed to be closely related to the mixed spin-singlet and triplet pairing, whose degree can be tuned by the strength of ASOC~\cite{Bauer2012,Smidman2017,Ghosh2020b}.  
For example, both Li$_2$Pt$_3$B and CaPtAs exhibits gap nodes~\cite{yuan2006,nishiyama2007,Shang2020}, an indication of superconductors with a 
spin-triplet pairing. 
More interestingly, TRS has been found to be broken in the superconducting state of CaPtAs~\cite{Shang2020}, thus confirming its unconventional SC. 
In both compounds, the ASOC-related band splitting is significant, i.e., $E_\mathrm{SOC}$/$k_\mathrm{B}$$T_c$ $\sim$ 800~\cite{Xie2019,Lee2005}. 
Our detailed $\mu$SR results suggest that \LBP\ exhibits a nodess SC with a preserved TRS, in hindsight compatible with its rather weak SOC. 
Upon replacing La with other nonmagnetic rare-earth metals
(such as Y, Lu, or Th), the Y$_4$Be$_{33}$Pt$_{16}$, Lu$_4$Be$_{33}$Pt$_{16}$,
and Th$_4$Be$_{33}$Pt$_{16}$ alloy compounds all undergo a superconducting transition at 
$T_c$ = 0.9, 0.7, and 1.0\,K, respectively~\cite{Svanidze2021,Amon2020,Kozelj2021}.
The $T_c$ of \LBP\ increases upon substituting La with Th, with the highest $T_c = 3.2$\,K being reached in (La$_{1-x}$Th$_x$)$_4$Be$_{22}$Pt$_{16}$ for
$x = 0.33$~\cite{Kozelj2021}.
Considering that all these materials share similar electronic band structures
with \LBP, they too are expected to be conventional superconductors.  
At the same time, there are also many NCSCs which, despite a significant $E_\mathrm{SOC}$, exhibit conventional superconducting properties, characterized by a fully-gapped SC with a preserved TRS~\cite{Shang2021b,Sharma2020,Shang2023,Tay2023,Ptok2019}.
Clearly, in NCSCs, the relationship between the unconventional
superconductivity and ASOC is not yet fully understood, and further
efforts, including the search for new NCSCs, are highly desirable.

\section{Summary} 
To conclude, the superconducting properties of the non\-cen\-tro\-sym\-metric
\LBP\ alloy were investigated by mag\-ne\-ti\-za\-tion-, heat-capacity-, and $\mu$SR measurements. 
The \LBP\ alloy is found to undergo a bulk superconducting transition at $T_c = 2.4$\,K.
The absence of additional muon-spin relaxation in the superconducting state confirms the lack of spontaneous magnetic fields below the onset of SC, suggesting a \emph{preserved} time-reversal symmetry 
in the superconducting state. 
The temperature dependence of the electronic specific heat and
of the superfluid density are well described by an 
$s$-wave model with a single gap, consistent with a \emph{fully-gapped} superconducting state in \LBP. 
Our results demonstrate that spin-singlet pairing dominates in \LBP\ and, 
hence, it behaves as a conventional superconductor, despite the
noncentrosymmetric crystal structure and the large spin-orbit coupling
of La and Pt. \tcr{Why yet another NCSC, expected to exhibit a complex
behaviour, shows only conventional features remains an open question.}

\vspace{5mm}
\noindent\textbf{Acknowledgements}\\
This work was supported by the Natural Science Foundation of Shanghai 
(Grants No.\ 21ZR1420500 and 21JC\-140\-2300), Natural Science Foundation
of Chongqing (Grant No.\ CSTB-2022NSCQ-MSX1678), National Natural Science Foundation of China (Grant No. 12374105) 
and the Schweizerische Nationalfonds zur F\"{o}r\-der\-ung der
Wis\-sen\-schaft\-lichen For\-schung (SNF) (Grant No.\ 200021\_169455).
E.S. is grateful for the support of the Christiane N\"{u}sslein-Volhard Stiftung.
We acknowledge the allocation of beam time at the Swiss muon source 
(Dolly {\textmu}SR spectrometer).

\vspace{5mm}
\noindent\textbf{Data availability statement}\\
All the data needed to evaluate the reported conclusions 
are presented in the paper. 
The $\mu$SR data were generated at the S$\mu$S  (Paul Scherrer Institut, Switzerland).
The \texttt{musrfit} software package is available online free of charge at 
http://lmu.web.psi.ch/musrfit/technical.

\vspace{5mm}
\noindent\textbf{References}\\
%
\bibliography{LaBePt}	

\providecommand{\newblock}{}
\begin{thebibliography}{10}
\expandafter\ifx\csname url\endcsname\relax
  \def\url#1{{\tt #1}}\fi
\expandafter\ifx\csname urlprefix\endcsname\relax\def\urlprefix{URL }\fi
\providecommand{\eprint}[2][]{\url{#2}}

\bibitem{Sigrist1991}
Sigrist M and Ueda K 1991 {\em Rev. Mod. Phys.\/} {\bf 63}(2) 239--311

\bibitem{Bauer2012}
Bauer E and Sigrist M (eds) 2012 {\em Non-{C}entrosymmetric
  {S}uperconductors\/} ({\em Lecture Notes in Physics\/} vol 847) (Berlin:
  Springer Verlag)

\bibitem{Yip2014}
Yip S 2014 {\em Annu. Rev. Condens. Matter Phys.\/} {\bf 5} 15--33

\bibitem{Bauer2004}
Bauer E, Hilscher G, Michor H, Paul C, Scheidt E~W, Gribanov A, Seropegin Y,
  No{\"e}l H, Sigrist M and Rogl P 2004 {\em Phys. Rev. Lett.\/} {\bf 92}
  027003

\bibitem{Smidman2017}
Smidman M, Salamon M~B, Yuan H~Q and Agterberg D~F 2017 {\em Rep. Prog.
  Phys.\/} {\bf 80} 036501

\bibitem{Ghosh2020b}
Ghosh S~K, Smidman M, Shang T, Annett J~F, Hillier A~D, Quintanilla J and Yuan
  H 2020 {\em J. Phys.: Condens. Matter\/} {\bf 33} 033001

\bibitem{Carnicom2018}
Carnicom E~M, Xie W, Klimczuk T, Lin J~J, G{\'o}rnicka K, Sobczak Z, Ong N~P
  and Cava R~J 2018 {\em Sci. Adv.\/} {\bf 4} eaar7969

\bibitem{Su2021}
Su H, Shang T, Du F, Chen C~F, Ye H~Q, Lu X, Cao C, Smidman M and Yuan H~Q 2021
  {\em Phys. Rev. Mater.\/} {\bf 5} 114802

\bibitem{Balakirev2015}
Balakirev F~F, Kong T, Jaime M, McDonald R~D, Mielke C~H, Gurevich A, Canfield
  P~C and Bud'ko S~L 2015 {\em Phys. Rev. B\/} {\bf 91}(22) 220505(R)

\bibitem{Bao2015}
Bao J~K, Liu J~Y, Ma C~W, Meng Z~H, Tang Z~T, Sun Y~L, Zhai H~F, Jiang H, Bai
  H, Feng C~M, Xu Z~A and Cao G~H 2015 {\em Phys. Rev. X\/} {\bf 5}(1) 011013

\bibitem{Tang2017}
Tang Z~T, Liu Y, Bao J~K, Xi C~Y, Pi L and Cao G~H 2017 {\em J. Phys.: Condens.
  Matter\/} {\bf 29} 424002

\bibitem{Tang2015}
Tang Z~T, Bao J~K, Wang Z, Bai H, Jiang H, Liu Y, Zhai H~F, Feng C~M, Xu Z~A
  and Cao G~H 2015 {\em Sci. China Mater.\/} {\bf 58} 16--20

\bibitem{yuan2006}
Yuan H~Q, Agterberg D~F, Hayashi N, Badica P, Vandervelde D, Togano K, Sigrist
  M and Salamon M~B 2006 {\em Phys. Rev. Lett.\/} {\bf 97} 017006

\bibitem{nishiyama2007}
Nishiyama M, Inada Y and Zheng G~q 2007 {\em Phys. Rev. Lett.\/} {\bf 98}
  047002

\bibitem{bonalde2005CePt3Si}
Bonalde I, Br{\"a}mer-Escamilla W and Bauer E 2005 {\em Phys. Rev. Lett.\/}
  {\bf 94} 207002

\bibitem{Shang2020}
Shang T, Smidman M, Wang A, Chang L~J, Baines C, Lee M~K, Nie Z~Y, Pang G~M,
  Xie W, Jiang W~B, Shi M, Medarde M, Shiroka T and Yuan H~Q 2020 {\em Phys.
  Rev. Lett.\/} {\bf 124} 207001

\bibitem{kuroiwa2008}
Kuroiwa S, Saura Y, Akimitsu J, Hiraishi M, Miyazaki M, Satoh K~H, Takeshita S
  and Kadono R 2008 {\em Phys. Rev. Lett.\/} {\bf 100} 097002

\bibitem{Hillier2009}
Hillier A~D, Quintanilla J and Cywinski R 2009 {\em Phys. Rev. Lett.\/} {\bf
  102} 117007

\bibitem{Barker2015}
Barker J~A~T, Singh D, Thamizhavel A, Hillier A~D, Lees M~R, Balakrishnan G,
  Paul D~M and Singh R~P 2015 {\em Phys. Rev. Lett.\/} {\bf 115} 267001

\bibitem{Shang2020b}
Shang T, Ghosh S~K, Zhao J~Z, Chang L~J, Baines C, Lee M~K, Gawryluk D~J, Shi
  M, Medarde M, Quintanilla J and Shiroka T 2020 {\em Phys. Rev. B\/} {\bf 102}
  020503(R)

\bibitem{Singh2014}
Singh R~P, Hillier A~D, Mazidian B, Quintanilla J, Annett J~F, Paul D~M,
  Balakrishnan G and Lees M~R 2014 {\em Phys. Rev. Lett.\/} {\bf 112} 107002

\bibitem{Shang2018a}
Shang T, Pang G~M, Baines C, Jiang W~B, Xie W, Wang A, Medarde M, Pomjakushina
  E, Shi M, Mesot J, Yuan H~Q and Shiroka T 2018 {\em Phys. Rev. B\/} {\bf 97}
  020502(R)

\bibitem{Shang2018b}
Shang T, Smidman M, Ghosh S~K, Baines C, Chang L~J, Gawryluk D~J, Barker J~A~T,
  Singh R~P, Paul D~M, Balakrishnan G, Pomjakushina E, Shi M, Medarde M,
  Hillier A~D, Yuan H~Q, Quintanilla J, Mesot J and Shiroka T 2018 {\em Phys.
  Rev. Lett.\/} {\bf 121} 257002

\bibitem{Shang2020ReMo}
Shang T, Baines C, Chang L~J, Gawryluk D~J, Pomjakushina E, Shi M, Medarde M
  and Shiroka T 2020 {\em npj Quantum Mater.\/} {\bf 5} 76

\bibitem{Shang2022b}
Shang T, Zhao J, Hu L~H, Ma J, Gawryluk D~J, Zhu X, Zhang H, Zhen Z, Yu B, Xu
  Y, Zhan Q, Pomjakushina E, Shi M and Shiroka T 2022 {\em Sci. Adv.\/} {\bf 8}
  eabq6589

\bibitem{Shang2022c}
Shang T, Ghosh S~K, Smidman M, Gawryluk D~J, Baines C, Wang A, Xie W, Chen Y,
  Ajeesh M~O, Nicklas M, Pomjakushina E, Medarde M, Shi M, Annett J~F, Yuan H,
  Quintanilla J and Shiroka T 2022 {\em npj Quantum Mater.\/} {\bf 7} 35

\bibitem{Shang2021a}
Shang T and Shiroka T 2021 {\em Front. Phys.\/} {\bf 9} 270

\bibitem{Sato2017}
Sato M and Ando Y 2017 {\em Rep. Prog. Phys.\/} {\bf 80} 076501

\bibitem{Qi2011}
Qi X~L and Zhang S~C 2011 {\em Rev. Mod. Phys.\/} {\bf 83} 1057--1110

\bibitem{Kallin2016}
Kallin C and Berlinsky J 2016 {\em Rep. Prog. Phys.\/} {\bf 79} 054502

\bibitem{Kim2018}
Kim H, Wang K, Nakajima Y, Hu R, Ziemak S, Syers P, Wang L, Hodovanets H,
  Denlinger J~D, Brydon P~M~R, Agterberg D~F, Tanatar M~A, Prozorov R and
  Paglione J 2018 {\em Sci. Adv.\/} {\bf 4} eaao4513

\bibitem{Ali2014}
Ali M~N, Gibson Q~D, Klimczuk T and Cava R~J 2014 {\em Phys. Rev. B\/} {\bf 89}
  020505(R)

\bibitem{Guan2016}
Guan S~Y, Chen P~J, Chu M~W, Sankar R, Chou F, Jeng H~T, Chang C~S and Chuang
  T~M 2016 {\em Sci. Adv.\/} {\bf 2} e1600894

\bibitem{Bian2016}
Bian G, Chang T~R, Sankar R, Xu S~Y, Zheng H, Neupert T, Chiu C~K, Huang S~M,
  Chang G, Belopolski I, Sanchez D~S, Neupane M, Alidoust N, Liu C, Wang B, Lee
  C~C, Jeng H~T, Zhang C, Yuan Z, Jia S, Bansil A, Chou F, Lin H and Hasan M~Z
  2016 {\em Nat. Commun.\/} {\bf 7} 10556

\bibitem{Sakano2015}
Sakano M, Okawa K, Kanou M, Sanjo H, Okuda T, Sasagawa T and Ishizaka K 2015
  {\em Nat. Commun.\/} {\bf 6} 8595

\bibitem{Sun2015}
Sun Z~X, Enayat M, Maldonado A, Lithgow C, Yelland E, Peets D~C, Yaresko A,
  Schnyder A~P and Wahl P 2015 {\em Nat. Commun.\/} {\bf 6} 6633

\bibitem{Neupane2016}
Neupane M, Alidoust N, Hosen M~M, Zhu J~X, Dimitri K, Xu S~Y, Dhakal N, Sankar
  R, Belopolski I, Sanchez D~S, Chang T~R, Jeng H~T, Miyamoto K, Okuda T, Lin
  H, Bansil A, Kaczorowski D, Chou F, Hasan M~Z and Durakiewicz T 2016 {\em
  Nat. Commun.\/} {\bf 7} 13315

\bibitem{Singh2020LaRh}
Singh D, Scheurer M~S, Hillier A~D, Adroja D~T and Singh R~P 2020 {\em Phys.
  Rev. B\/} {\bf 102}(13) 134511

\bibitem{Arushi2021}
Arushi, Singh D, Hillier A~D, Scheurer M~S and Singh R~P 2021 {\em Phys. Rev.
  B\/} {\bf 103}(17) 174502

\bibitem{Mondal2012}
Mondal M, Joshi B, Kumar S, Kamlapure A, Ganguli S~C, Thamizhavel A, Mandal
  S~S, Ramakrishnan S and Raychaudhuri P 2012 {\em Phys. Rev. B\/} {\bf 86}(9)
  094520

\bibitem{Xu2020}
Xu X, Li Y and Chien C~L 2020 {\em Phys. Rev. Lett.\/} {\bf 124}(16) 167001

\bibitem{Ramakrishnan2017}
Ramakrishnan S, Joshi B and Thamizhavel A 2017 {\em Philos. Mag.\/} {\bf 97}
  3460--3476

\bibitem{Parzyk2014}
Parzyk N~A 2014 {\em Muon and neutron studies of unconventional
  superconductors\/} Ph.D. thesis University of Warwick Coventry

\bibitem{Xie2019}
Xie W, Zhang P~R, Shen B, Jiang W~B, Pang G~M, Shang T, Gao C, Smidman M and
  Yuan H~Q 2020 {\em Sci. China-Phys. Mech. Astron.\/} {\bf 63} 237412

\bibitem{Shang2021b}
Shang T, Xie W, Zhao J~Z, Chen Y, Gawryluk D~J, Medarde M, Shi M, Yuan H~Q,
  Pomjakushina E and Shiroka T 2021 {\em Phys. Rev. B\/} {\bf 103} 184517

\bibitem{Sharma2020}
Sharma S, Arushi, Motla K, Beare J, Nugent M, Pula M, Munsie T~J, Hillier A~D,
  Singh R~P and Luke G~M 2021 {\em Phys. Rev. B\/} {\bf 103} 104507

\bibitem{Shang2023}
Shang T, Zhao J~Z, Hu L~H, Gawryluk D~J, Zhu X~Y, Zhang H, Meng J, Zhen Z~X, Yu
  B~C, Zhou Z, Xu Y, Zhan Q~F, Pomjakushina E and Shiroka T 2023 {\em Phys.
  Rev. B\/} {\bf 107}(22) 224504

\bibitem{Tay2023}
Tay D, Shang T, Rosa P~F~S, Santos F~B, Thompson J~D, Fisk Z, Ott H~R and
  Shiroka T 2023 {\em Phys. Rev. B\/} {\bf 107}(6) 064507

\bibitem{Ptok2019}
Ptok A, Domieracki K, Kapcia K~J, \L{}a\.{z}ewski J, Jochym P~T, Sternik M,
  Piekarz P and Kaczorowski D 2019 {\em Phys. Rev. B\/} {\bf 100}(16) 165130

\bibitem{Svanidze2021}
Svanidze E, Amon A, Nicklas M, Prots Y, Juckel M, Rosner H, Burkhardt U, Avdeev
  M, Grin Y and Leithe-Jasper A 2021 {\em Phys. Rev. Mater.\/} {\bf 5}(7)
  074801

\bibitem{Amon2020}
Amon A, Svanidze E, Prots Y, Nicklas M, Burkhardt U, Ormeci A, Leithe-Jasper A
  and Grin Y 2020 {\em Dalton Trans.\/} {\bf 49} 9362--9368

\bibitem{Kozelj2021}
Ko\v{z}elj P, Juckel M, Amon A, Prots Y, Ormeci A, Burkhardt U, Brando M,
  Leithe-Jasper A, Grin Y and Svanidze E 2021 {\em Sci. Rep.\/} {\bf 11} 22352

\bibitem{Suter2012}
Suter A and Wojek B~M 2012 {\em Phys. Procedia\/} {\bf 30} 69--73

\bibitem{Joseph1972}
Joseph R~R, Gschneidner K~A and Hungsberg R~E 1972 {\em Phys. Rev. B\/} {\bf
  5}(5) 1878--1885

\bibitem{Holt1981}
Holt B, Ramsden J, Sample H and Huber J 1981 {\em Physica B+C\/} {\bf 107}
  255--256

\bibitem{Yaouanc2011}
Yaouanc A and de~R\'eotier P~D 2011 {\em Muon Spin Rotation, Relaxation, and
  Resonance: Applications to Condensed Matter\/} (Oxford: Oxford University
  Press)

\bibitem{Maisuradze2009}
Maisuradze A, Khasanov R, Shengelaya A and Keller H 2009 {\em J. Phys.:
  Condens. Matter\/} {\bf 21}(7) 075701 and references therein

\bibitem{Barford1988}
Barford W and Gunn J~M~F 1988 {\em Physica C\/} {\bf 156} 515--522

\bibitem{Brandt2003}
Brandt E~H 2003 {\em Phys. Rev. B\/} {\bf 68} 054506

\bibitem{Tinkham1996}
Tinkham M 1996 {\em Introduction to Superconductivity\/} 2nd ed (Mineola, NY:
  Dover Publications)

\bibitem{Carrington2003}
Carrington A and Manzano F 2003 {\em Physica C\/} {\bf 385} 205--214

\bibitem{Volovik1993}
Volovik G~E 1993 {\em JETP Lett.\/} {\bf 58}(6) 469--473

\bibitem{Wen2004}
Wen H~H, Liu Z~Y, Zhou F, Xiong J, Ti W, Xiang T, Komiya S, Sun X and Ando Y
  2004 {\em Phys. Rev. B\/} {\bf 70}(21) 214505

\bibitem{Chen2017}
Chen J~T, Sun Y, Yamada T, Pyon S and Tamegai T 2017 {\em J. Phys. Conf.
  Ser.\/} {\bf 871} 012016

\bibitem{Bouquet2001a}
Bouquet F, Fisher R~A, Phillips N~E, Hinks D~G and Jorgensen J~D 2001 {\em
  Phys. Rev. Lett.\/} {\bf 87}(4) 047001

\bibitem{Huang2007}
Huang C~L, Lin J~Y, Chang Y~T, Sun C~P, Shen H~Y, Chou C~C, Berger H, Lee T~K
  and Yang H~D 2007 {\em Phys. Rev. B\/} {\bf 76}(21) 212504

\bibitem{Caroli1964}
Caroli C, De~Gennes P~G and Matricon J 1964 {\em Phys. Lett.\/} {\bf 9} 307

\bibitem{Shang2020NbIr}
Xu Y, J\"ohr S, Das L, Kitagawa J, Medarde M, Shiroka T, Chang J and Shang T
  2020 {\em Phys. Rev. B\/} {\bf 101} 134513

\bibitem{Kubo1967}
Kubo R and Toyabe T 1967 A stochastic model for low-field resonance and
  relaxation {\em Magnetic Resonance and Relaxation. Proceedings of the {XIV}th
  {C}olloque {A}mp\`ere\/} ed Blinc R (Amsterdam: North-Holland) pp 810--823

\bibitem{Lee2005}
Lee K~W and Pickett W~E 2005 {\em Phys. Rev. B\/} {\bf 72}(17) 174505

\end{thebibliography}
	
\end{document}